\documentclass[twocolumn,aps,pre,amsmath,showpacs,superscriptaddress]{revtex4-1}
\usepackage{graphicx,dcolumn,bm,amssymb,amsmath,ulem,indentfirst,amsthm}
\usepackage[toc,page]{appendix}
\usepackage{color}

\theoremstyle{plain}
\def\be{\begin{equation}}
\def\ee{\end{equation}}

\newtheorem*{theorem*}{Theorem}

\begin{document}
\author{Bingyu Cui}
\affiliation{Statistical Physics Group, Department of Chemical
Engineering and Biotechnology, University of Cambridge, Philippa Fawcett Drive,
CB3
0AS Cambridge, U.K.}
\author{Zach Evenson}
 \affiliation{Heinz Maier-Leibnitz Zentrum (MLZ) and Physik Department,
Technische
Universit\"{a}t M\"{u}nchen, Lichtenbergstrasse 1, 85748 Garching, Germany}
\author{Beibei Fan}
\affiliation{Department of Physics, Renmin University of China, Beijing 100872,
China}
\author{Mao-Zhi Li}
\affiliation{Department of Physics, Renmin University of China, Beijing 100872,
China}
\author{Wei-Hua Wang}
\email{whw@iphy.ac.cn}
\affiliation{Institute of Physics, Chinese Academy of Sciences, Beijing 100190,
China}
\author{Alessio Zaccone}
\email{az302@cam.ac.uk}
\affiliation{Statistical Physics Group, Department of Chemical
Engineering and Biotechnology, University of Cambridge, Philippa Fawcett Drive,
CB3
0AS Cambridge, U.K.}
\affiliation{Cavendish Laboratory, University of Cambridge, JJ Thomson
Avenue, CB3 0HE Cambridge,
U.K.}

\begin{abstract}
We employ an atomic-scale theory within the framework of nonaffine lattice
dynamics to uncover the origin of the Johari-Goldstein (JG) $\beta$-relaxation
in metallic glasses (MGs). Combining simulation and experimental data with our
theoretical approach, we reveal that the large mass asymmetry between the
elements in a La$_{60}$Ni$_{15}$Al$_{25}$ MG leads to a clear separation in the
respective relaxation time scales, giving strong evidence that JG relaxation is
controlled by the lightest atomic species present. Moreover, we show that only
qualitative features of the vibrational density of states determine the overall
observed mechanical response of the glass, paving the way for a possible
unified theory of secondary relaxations in glasses.
\end{abstract}

\pacs{}
\title{Possible origin of $\beta$-relaxation in amorphous metal alloys from
atomic-mass differences of the constituents}
\maketitle

\section{Introduction}
The diversity of atomic motion in metallic glasses (MGs) is central to their
unique physical and mechanical properties. The primary or $\alpha$-relaxation
underlies the drastic slowing down of the collective atomic dynamics during the
transition from a viscous supercooled liquid to a glassy solid upon cooling,
and its origin is still an outstanding problem in condensed matter physics.
Indeed, like many other disordered solids, such as polymers and molecular
glasses, MGs exhibit an entire class of secondary relaxations that persist even
well below the glass transition temperature
$T_g$~\cite{Yu2014a,Yu2013b,Ku2017}. These phenomena are broadly referred to as
$\beta$-relaxations and occur on time scales much shorter than that of the
$\alpha$-relaxation. The Johari-Goldstein (JG) $\beta$-relaxation is the most
well known amongst these, due to its ubiquity in all types of
glasses~\cite{Jo1970,Ng2000}. Although the exact atomic-scale mechanism
underlying the JG $\beta$-relaxation in MGs is still not clear, there appears
to be a correlation to the $\alpha$-relaxation, deformation and mechanical
properties (see ~\cite{Yu2014a} and references therein). In this regard,
unraveling the atomic-scale dynamical features of the JG $\beta$-relaxation
would represent considerable progress in our current understanding of its
microscopic origin and its impact on the physical and materials properties of
glasses~\cite{Ruta2017}.

A key open question is about the role of different atomic/molecular
constituents
in the various relaxation processes, and in particular whether a relaxation
process is
controlled by the dynamics of a particular type of constituent(s). In the case
of organic molecular glasses
it has been recently argued that all molecules seem to participate in the JG
relaxation, although not all at once ~\cite{Cicerone2017}.
This problem has not been investigated in metallic glasses, although the
relative contributions of different atomic species
to the peak temperature of the JG relaxation has been addressed in
~\cite{Zhu2014}.

While many studies have examined both the structural and relaxational features
of the JG $\beta$-relaxation in
MGs~\cite{Evenson2014b,Yu2017,Wang2015a,Liu2014} the connection to the
atomic-scale vibrational properties remains to date greatly unexplored. The JG
$\beta$-relaxation in MGs generally occurs on microsecond time scales, some
several orders of magnitude smaller than the $\alpha$-relaxation of the
glass~\cite{Yu2017, Liu2017}. However, accessing the atomic-scale dynamics of
MGs in this temporal regime is both experimentally and computationally
challenging. Novel coherent x-ray scattering techniques probe collective atomic
motion on time scales larger than about one second~\cite{Wang2015a, Ruta},
while molecular dynamics (MD) simulations of the MG glassy-state dynamics have
been only recently successfully tested up to 10 microseconds~\cite{Yu2017}.

Here, we combine experimental and simulation investigations with a microscopic
theoretical framework of viscoelastic response and relaxation of MGs. With this
novel approach, we are able to unveil the atomic-scale dynamics in MGs on
time-scales over some 12 orders of magnitude, thus providing necessary,
complementary information for advanced simulation and experimental studies.

Considering the success of our recent theoretical work in linking the
low-energy boson peak (BP) with $\alpha$-relaxation and dynamical heterogeneity
in glasses~\cite{Cui,Cui2}, the results presented in this paper give new
insight into the atomic-scale dynamical facets of the JG $\beta$-relaxation in
MGs. In particular, we are able to show strong evidence that the JG
$\beta$-relaxation is controlled by the smallest (lightest) atomic scale
species present in the MG, and that the existence of two relaxation modes
~($\alpha$ and JG $\beta$) can be traced down to the large differences in
atomic mass of the metallic elements that comprise the MG.

\section{Experimental methods}
\subsection{Dynamical mechanical analysis}
The dynamical mechanical analysis (DMA) experiments we carried out according to
the procedure outlined in Ref.~\cite{Zhu2014} using a TAQ800 dynamical
mechanical analyzer. Fully amorphous cylindrical samples of
La$_{60}$Ni$_{15}$Al$_{25}$ with a diameter of 2 mm were tested using the
single-cantilever bending method in an isothermal mode with a strain amplitude
of 5 $\mu$m, temperature step of 3 K and discrete testing frequencies of 1, 2,
4, 8, and 16 Hz. The complex viscoelastic shear modulus is obtained as
$G(\omega,T)=G'(\omega,T)+iG''(\omega,T)$ as a function of test frequency
$\omega$ and temperature $T$, with mechanical relaxations appearing peaks in
the loss modulus $G''(\omega,T)$.

\subsection{Inelastic neutron scattering}
Glassy ribbons of La$_{60}$Ni$_{15}$Al$_{25}$ were produced by melt spinning at
the Institute for Physics, Chinese Academy of Sciences in Beijing. About 12 m
of ribbons with a cross-section of 2.5 $\times$ 0.06 mm$^2$ were placed in a
thin-walled aluminum hollow cylinder (height 51 mm, diameter 20 mm, thickness
0.55 mm) for the inelastic neutron scattering (INS) experiments at the
time-of-flight spectrometer TOFTOF in Garching. An incident wavelength of
$\lambda_i = 2.8$\,\AA\, resulted in an accessible momentum transfer range of
$0.8 \leq q \leq 4.2$ \AA$^{-1}$ at zero-energy transfer. The raw data were
normalized to a vanadium standard, corrected for empty container scattering and
sample shelf-absorption, and interpolated to constant $q$ in order to obtain
the dynamic structure factor. The background was corrected by separate
measurements of the cryostat with an empty sample holder. As the scattering
probability of the ribbons was calculated to be around 8\,\%, multiple
scattering effects were neglected.

In order to access the largest energy transfer range available, only the data
located on the neutron energy gain side of the spectrometer were analyzed. In a
multi-component system with predominantly coherent scatterers, a generalized,
neutron-weighted vibrational density of states (VDOS) $D(\omega_p)$ can be
obtained under the incoherent one-phonon approximation, where the measured
dynamic structure factor, integrated over the accessible $q$-range, is
proportional to $D(\omega_p)/\omega_p^2$~\cite{Meyer1996}. The neutron-weighted
VDOS was obtained in an iterative procedure using the FRIDA-1
software~\cite{frida,Wuttke1993}.

\section{Molecular Dynamics simulations}
Classical molecular dynamics (MD) simulations were performed for the
La$_{60}$Ni$_{15}$Al$_{25}$ metallic alloy system using the LAMMPS
package~\cite{Pl1995}. The interatomic interactions were described by the
embedded-atom method (EAM) potential in Ref.~\cite{Sh2008}. Details can be
found in the Appendix A.
To obtain the VDOS $D(\omega_p)$ of the system at various temperatures, the
direct diagonalization method was adopted, in which the steepest-descent method
is carried out for the final configuration.

The structure model contains 10,000 atoms in a cubic box with periodic boundary
conditions applied in three dimensions. It was first fully equilibrated at
T=2000 K for 1 ns in the NPT (isobaric and isothermal) ensemble, then cooled
down to 300 K with a cooling rate of 10$^{12}$ K/s. In the cooling process, the
box size was adjusted to give zero pressure. At 300 K, the structure was then
relaxed for 2 ns in the NPT ensemble. To obtain the atomic structures at 330,
360, 390, and 410 K, the structure at 300 K was then heated with heating rate
of 10$^{10}$ K/s, and then relaxed for 2 ns in NPT ensemble at each temperature
of interest. The MD step was set to be 2 fs.

The dynamical matrix corresponding to the potential energy minimum reached by
LAMMPS line search algorithms minimization is given by
\begin{equation}
H_{ij} = \frac{1}{\sqrt{m_{i}m_{j}}} \frac{\partial^{2} U}{\partial
\underline{x}_i
\partial \underline{x}_j}
\end{equation}
where $U$ is the total internal energy of the system (which is a function of
all atoms' coordinates), $m_{i}$ is the mass of atom $i$ and $\underline{x}_i$
is the coordinate vector of atom $i$. The VDOS can be calculated by directly
diagonalizing the dynamical matrix as
\begin{equation}
D(\omega_p) = \frac{1}{3N-3} \sum_{\lambda} \delta (\omega_p -
\omega_{\lambda}),
\end{equation}
where $\omega_{\lambda}$ is the eigenfrequency.

\section{Nonaffine lattice dynamics}
\subsection{From the Generalized Langevin Equation to the dynamic viscoelastic
moduli}
The dynamics of atoms in disordered solids is typically nonaffine, which means
that the atoms in the deformed configuration do not sit in the positions
prescribed by the strain tensor, i.e. they do not get displaced according to an
affine transformation. The latter would give the new position of the atom from
the left-multiplication of strain tensor and position vector of the atom at
rest. Instead, in disordered systems, the atom in the affine position receives
forces from the nearest-neighbours which do not balance (they would balance and
cancel to zero in a centrosymmetric crystal, owing to local inversion symmetry
of the lattice). Hence lattice dynamics for amorphous materials has to be
rewritten to take these facts into account~\cite{Lemaitre} which eventually
leads to softening of the elastic constants~\cite{Zaccone2011} and new physics
which is currently being explored.

Upon applying a deformation described by the strain tensor
$\underline{\underline{\eta}}$, the dynamics of a tagged particle $i$
interacting with other atoms in the reference frame satisfies the following
equation for the (mass-scaled) displacement
$\{\underline{x}_i(t)=\underline{\mathring{q}}_i(t)-\underline{\mathring{q}}_i\}$
around a known rest frame $\underline{\mathring{q}}_i$ (see Ref.~\cite{Cui} for
derivation):
\begin{equation}
\label{eq:NAD1}
\frac{d^2\underline{x}_i}{dt^2}+\int_{-\infty}^{t}\nu_i(t-t')\frac{d\underline{x}_i}{dt'}dt'+\sum_{j}\underline{\underline{H}}_{ij}\underline{x}_j=\underline{\Xi}_{i,xy}\eta_{xy}.
\end{equation}
Note that the summation convention over repeated indices is not used. This
equation can be solved by performing Fourier transformation followed by normal
mode decomposition that decomposes the 3N-vector $\tilde{\underline{x}}$ (that
contains positions of all atoms) into normal modes
$\tilde{\underline{x}}=\hat{\tilde{x}}_p(\omega)\underline{\phi}_p$ ($p$ is the
index labeling the normal modes).  Note that we specialize on time-dependent
shear
strain $\eta_{xy}(t)$. {For this case, the vector
$\underline{\Xi}_{i,xy}$ represents the force per unit strain acting on atom
$i$ due to the motion of its nearest-neighbors (see e.g.~\cite{Lemaitre} for a
more detailed discussion).

As shown in the Appendix A, Eq. (3) can be manipulated into the
following form:
\begin{equation}
-\omega^2(\underline{\underline{\Phi}}^{T}\cdot\tilde{\underline{x}})+i\omega\underline{\underline{\Phi}}^T\tilde{\underline{\underline{\nu}}}(\omega)\underline{\underline{\Phi}}\underline{\underline{\Phi}}^{T}\cdot\tilde{\underline{x}}
+\underline{\underline{D}}~(\underline{\underline{\Phi}}^{T}\cdot\underline{\tilde{x}})
=\underline{\underline{\Phi}}^{T}\cdot\underline{\Xi}_{xy}\tilde{\eta}_{xy},
\end{equation}
where the matrix $\underline{\underline{\Phi}}$ consists of the $3N$
eigenvectors $\underline{\phi}_p$ of the Hessian.
Here, we have
$(\underline{\underline{\Phi}}^{T}\underline{\underline{\tilde{\nu}}}\underline{\underline{\Phi}})_{mn}=\sum_i\Phi_{im}\Phi_{in}\tilde{\nu}_i$
and $(\Phi^T\Phi)_{mn}=\sum_i\Phi_{im}\Phi_{in}=\delta_{mn}$ where
$\underline{\underline{\tilde{\nu}}}$ is the diagonal matrix made  by
$\tilde{\nu}_i(\omega)$ along the diagonal.
%In principle, one would have different memory kernels $\tilde{\nu}_i(\omega)$
for different tagged particles $i$ and in general, one cannot find a solution
without simplifying the term
$\underline{\underline{\Phi}}^T\underline{\underline{\nu}}
\underline{\underline{\Phi}}$,
which establishes coupling between different eigenmodes contributions to the
friction.

The friction term coupled to the \textit{p}-th normal mode is thus
$i\omega\sum_{im}\Phi_{im}\Phi_{ip}\tilde{\nu}_i$. At this point of the
analysis, we need to work with the assumption that
$\underline{\underline{\Phi}}^T\underline{\underline{\nu}}
\underline{\underline{\Phi}}$
is a diagonal matrix. In physical terms, this means that the damping is not
correlated across different eigenmodes. This is an approximation used within
this framework to make the model solvable~\cite{Cui}. Thus, the friction that
the \textit{p}-th mode feels is dominated by
$i\omega\sum_{i}(\Phi_{ip})^2\tilde{\nu}_i$. This result is used in the
section below to justify the form of memory kernel for the
friction coefficient based on differences in atomic mass of the
constituents.

As derived in our previous work~\cite{Cui}, we use the GLE Eq. (3) under normal
mode decomposition while accounting for nonaffine displacements to derive a
microscopic expression for the complex viscoelastic modulus
\begin{equation}
\label{eq:GLE}
G^*(\omega)=G_A-3\rho\int_0^{\omega_{D}}\frac{D(\omega_p)\Gamma(\omega_p)}{\omega_p^2-\omega^2+i\tilde{\nu}(\omega)\omega}d\omega_p
\end{equation}
where we have dropped the Cartesian indices for convenience and $\rho=N/V$
denotes the atomic density of the solid. $\Gamma(\omega_p)$ is a function which
describes the correlation of nonaffine forces in the frequency
shell~\cite{Lemaitre, Zaccone2011, Milkus_viscoelastic}.

\subsection{Qualitative arguments for the form of friction kernel in
La$_{60}$Ni$_{15}$Al$_{25}$}
As has been shown above in the context of Eq. (4), the friction that the
p-th mode feels is given by $\sum_i(\Phi_{ip})^2\nu_i$. We expand this term
explicitly in terms of the different atomic species which form the alloy:
\begin{align}
\sum_i(\Phi_{ip})^2\nu_i&\sim\sum_{Al}^{25}(\Phi_{ip}^2)\sum_\alpha\frac{m_\alpha}{m_{Al}}\frac{c_{\alpha,Al}^2}{\omega_\alpha^2}\cos{(\omega_\alpha
t)}\notag\\
&+\sum_{La}^{60}(\Phi_{ip}^2)\sum_\alpha\frac{m_\alpha}{m_{La}}\frac{c_{\alpha,La}^2}{\omega_\alpha^2}\cos{(\omega_\alpha
t)}\notag\\
&+\sum_{Ni}^{15}(\Phi_{ip}^2)\sum_\alpha\frac{m_\alpha}{m_{Ni}}\frac{c_{\alpha,Ni}^2}{\omega_\alpha^2}\cos{(\omega_\alpha
t)}\notag\\
&=\sum_\alpha\sum_{Al}^{25}(\Phi_{ip})^2\frac{m_\alpha}{m_{Al}}\frac{c^2_{\alpha,Al}}{\omega_\alpha^2}\cos{(\omega_\alpha
t)}\notag\\
&+\sum_\alpha\sum_{La}^{60}(\Phi_{ip})^2\frac{m_\alpha}{m_{La}}\frac{c^2_{\alpha,La}}{\omega_\alpha^2}\cos{(\omega_\alpha
t)}\notag\\
&+\sum_\alpha\sum_{Ni}^{15}(\Phi_{ip})^2\frac{m_\alpha}{m_{Ni}}\frac{c^2_{\alpha,Ni}}{\omega_\alpha^2}\cos{(\omega_\alpha
t)}\notag\\
\end{align}

The role of $\Phi_{ip}$ here is to give a weight to each $\nu_i$ contribution
in the sum. All these sums could be written also as integrals upon replacing
the discrete variable $\omega_{\alpha}$ with the continuous eigenfrequency
$\omega_p$ and introducing the VDOS as a factor in the integral over
$\omega_p$. Here, one can find that each term is inversely proportional to the
mass of the atomic species in question. We note that the atomic mass of La
(138.9\,u) is more than twice as large as the mass of Ni (58.7\,u) and five
times larger than the mass of Al (26.98\,u), which gives a much larger weight
in the sum to the Al and Ni terms. Hence, taking also stoichiometry into
account, the two terms relative to Ni and Al considered together are about
three times larger than the contribution of the La term.

In order to strengthen this claim, we also consider the role of the unknown
dynamical coupling coefficients $c_\alpha$ which appear in Eq. (6).
While the values of these coefficients cannot be determined from first
principles, we can still obtain valuable indications about the probable
magnitude by considering quantities like the partial $g(r)$ functions in the
system. Since these coefficients are associated with medium-range (or
generically, beyond-short-range) dynamics, features in the $g(r)$ may give an
indication about relative magnitude of dynamical coupling between different
species in the alloy.

Also, while $g(r)$ is a static structural quantity, it is also true that it is
directly related
to dynamics via the Boltzmann inversion relation which yields the potential of
mean force as
$V_{mfp}/k_B T= -\ln g(r)$. In turn, the potential of mean force represents the
interaction energy between two atoms mediated by the presence of all other
atoms in the system, hence it also contains many-body effects. Therefore,
$g(r)$ is directly related to the potential of mean force which in turn
influences the correlated motions (hence the dynamics) of the atoms and
establishes (e.g. through long-range attractions) the dynamic coupling.

Consideration of the pair correlation function obtained from simulations and
shown in Fig. (1) indicates that there is a clear broad peak
for Al-Al in the regime of the medium-range order. This supports our claim that
the JG $\beta$-relaxation is due to medium-range correlations and coupling
between Al atoms. This broad peak of Al-Al with respect to the short-range
order peak stands out in comparison with the other contributions in the
medium-range regime.

Finally, not only the pre-factor of the memory function of La will be smaller
compared to the other two atomic species, for the reason above, but also the
characteristic time-scale of memory decay associated with La will be 
comparatively larger, as the relaxation time is typically
inversely proportional to the mass (or at least inversely proportional to
square root of the mass). Hence, the contribution of La to memory and, hence,
to the intermediate scattering function (ISF) would be at a somewhat longer
time-scales compared to Ni. Additionally, this contribution would be probably
hybridized or obscured by Ni, which has a larger prefactor and would explain
why we result in only two decays in our model for the ISF and memory function.

These arguments, which indicate that the La-term in the form of
the memory function given by Eq. (6) may be negligible, can be summarized
as follows: (i) the mass-factor in the denominator makes the contribution of
La about three times smaller than the two contributions of Ni and Al taken
together; (ii) the main medium-range contributions to the features of the
$g(r)$ emanate from Al, which corroborates the hypothesis that the $c_\alpha$
coefficients are larger for Al and justify dominance of Al dynamics in the JG
$\beta$-relaxation; (iii) if modeled as a third stretched-exponential function,
the contribution of La would have a larger characteristic time-scale of decay
and would show up at longer times, probably masked or hybridized with the Ni
contribution.
Based on this approximation, the form of memory function for the interatomic
friction in Eq. (6) reduces to

\begin{align}
\nu(t)&=\sum_\alpha\sum_{Ni}^{15}(\Phi_{ip})^2\frac{m_\alpha}{m_{Ni}}\frac{c^2_{\alpha,Ni}}{\omega_\alpha^2}\cos{(\omega_\alpha
t)}\notag\\
&+\sum_\alpha\sum_{Al}^{25}(\Phi_{ip})^2\frac{m_\alpha}{m_{Al}}\frac{c^2_{\alpha,Al}}{\omega_\alpha^2}\cos{(\omega_\alpha
t)}\notag\\
&=\nu_{1}(t)+\nu_{2}(t).
\end{align}
where $\nu_1(t)$ and $\nu_2(t)$ are two generic functions of time that will be
specified in the next section.

\section{Relation between friction memory kernel and intermediate scattering
function}
For a supercooled liquid, a relationship between the time-dependent friction,
which is dominated by slow collective dynamics, and the intermediate scattering
function
has been famously derived within kinetic theory by Sjoegren and
Sjoelander~\cite{Sjoegren} (see also Ref.\cite{Bagchi}):
\begin{equation}
\label{eq:Sj}
\nu(t)=\frac{\rho k_{B}T}{6\pi^2 m}\int_{0}^{\infty}dq q^{4} F_{s}(q,t)c(q)^{2}
F(q,t)
\end{equation}
where $m$ is a characteristic mass, $c(q)$ is the direct correlation function
of liquid-state theory, $F(q,t)$ is the intermediate scattering function, and
$F_{s}(q,t)$ is the self-part of $F(q,t)$~\cite{Sjoegren}. All of these
quantities are functions of the wave-vector $q$ and the integral over $q$
leaves a time-dependence of
$\nu(t)$, which is exclusively given by the product $F_{s}(q,t)F(q,t)$. Upon
further approximating $F_{s}(q,t)F(q,t)\sim F(q,t)^{2}$, we obtain an
intermediate scattering function via

\begin{equation}
\label{eq:Fqt}
F(q,t) \sim \sqrt{\nu(t)}.
\end{equation}

That the VDOS is related to $\nu(t)$ becomes evident upon considering the
following relation, which holds for the particle-bath Hamiltonian from which
Eq.~\ref{eq:NAD1} is derived~\cite{Cui,Cui2,Zwanzig}
\begin{equation}
\nu(t)=\int_0^{\infty}d\omega_p
D(\omega_p)\frac{\gamma(\omega_p)^2}{\omega_p^2}\cos{\omega_p t},
\end{equation}
where $\gamma(\omega_p)$ is the continuous spectrum of coupling constants
which couple the dynamics of the tagged atom to that of all the oscillators
forming the bath, which represent all the other atomic degrees of freedom in
the material.

\begin{figure}
\begin{center}
\includegraphics[width=\columnwidth]{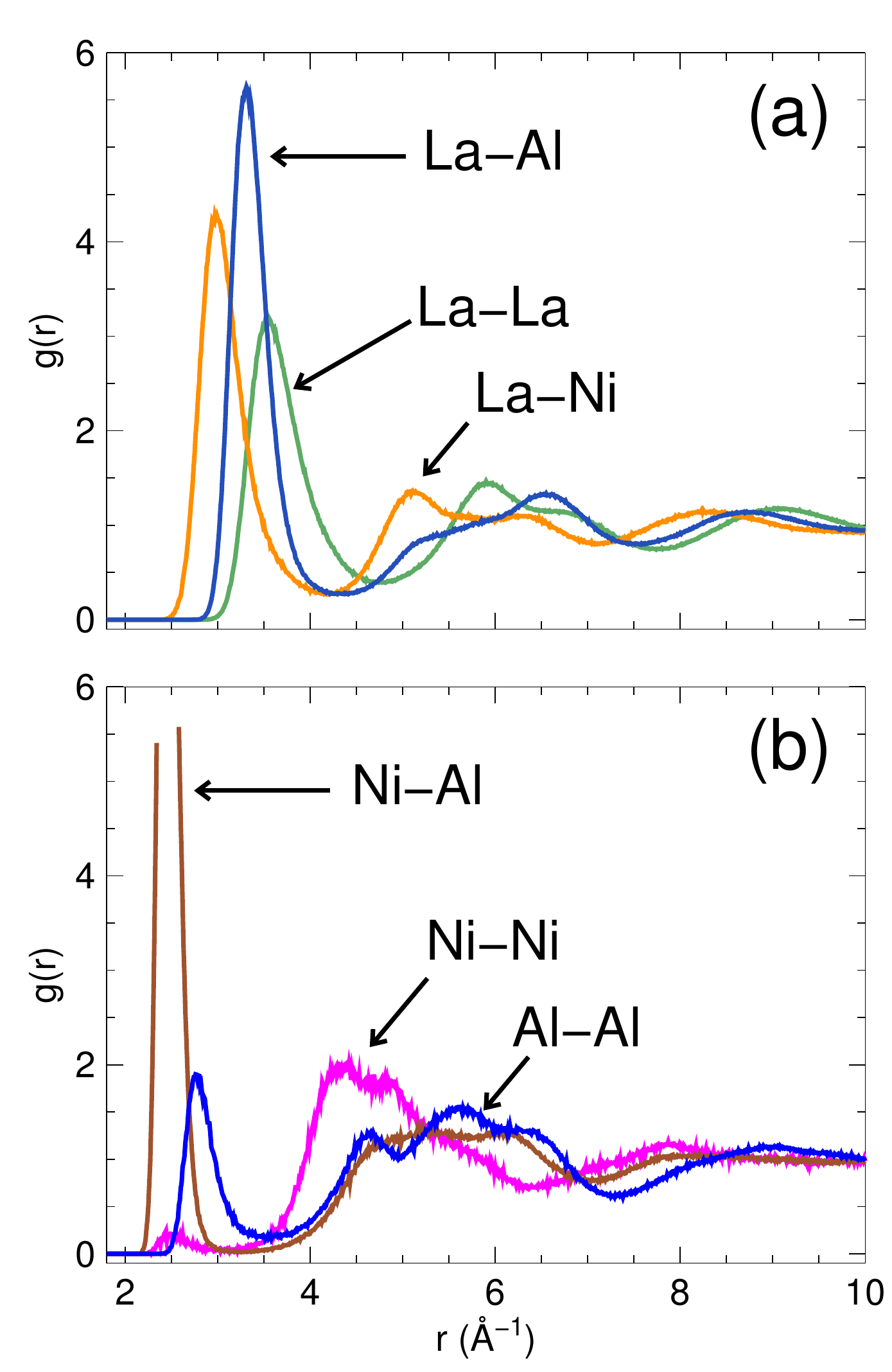}
\caption{\label{fig:rdf} Partial contributions to the radial distribution
function $g(r)$, as calculated from MD simulations for
\textrm{La}$_{60}$\textrm{Ni}$_{15}$\textrm{Al}$_{25}$ at $T=300$ K. The
large maximum of the Ni-Al partial in (b) occurs at $g(r_{\textrm{max}}) = 12$,
which falls out of the range of the vertical axis of the plot.}
\end{center}
\end{figure}

\begin{figure}
\begin{center}
\includegraphics[width=\columnwidth]{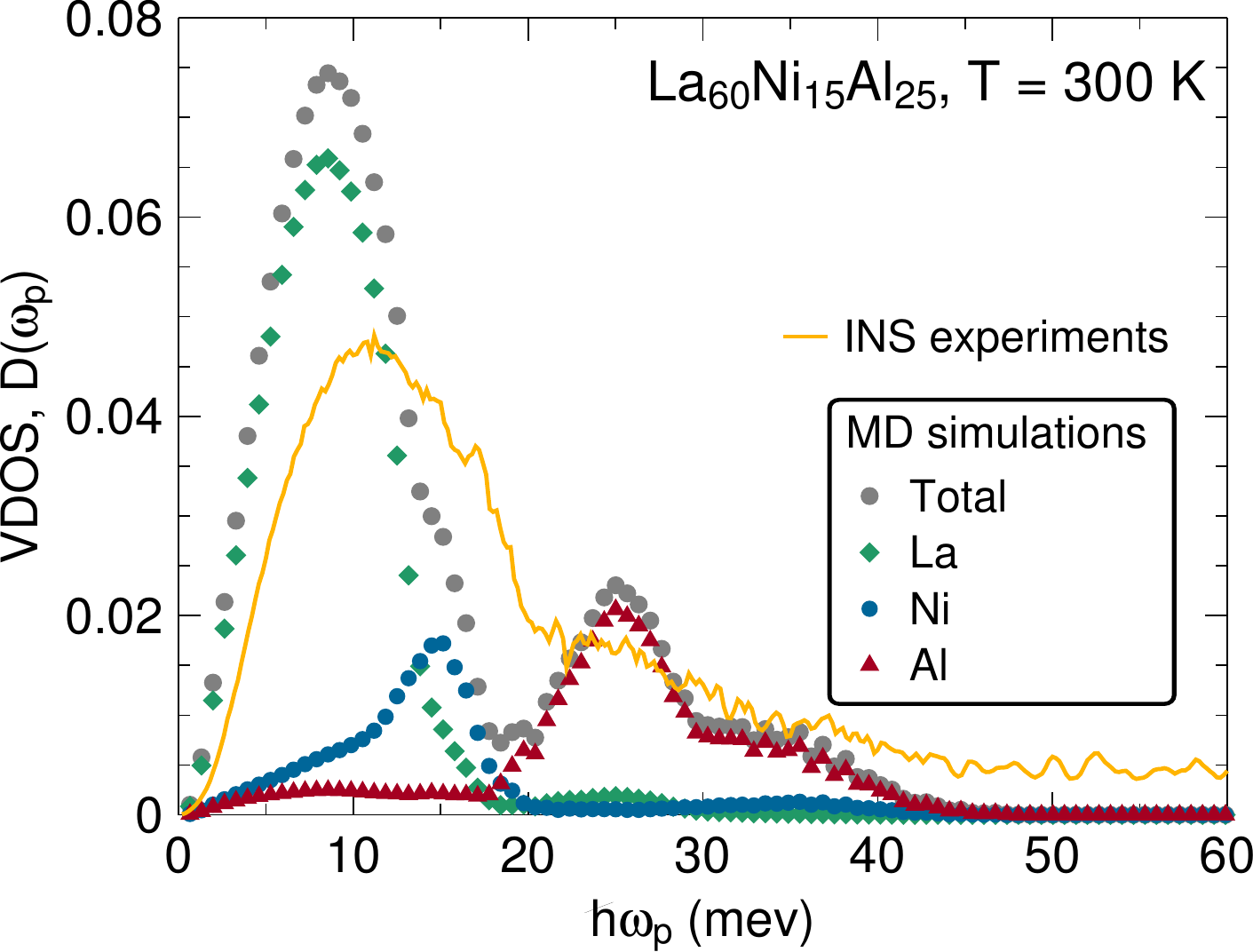}
\caption{\label{fig:DOS} Vibrational density of states (VDOS) of
\textrm{La}$_{60}$\textrm{Ni}$_{15}$\textrm{Al}$_{25}$ at $T=300$ K as
determined in INS experiments (solid line) and MD simulations (symbols).}
\end{center}
\end{figure}

\section{Results and discussion}
\subsection{Radial distribution function and partials thereof}
From the MD simulations we obtain the partial pair correlation functions $g(r)$
for all atomic pairs and show these in Fig.~\ref{fig:rdf}. The partial
functions shown in Fig.~\ref{fig:rdf}b clearly indicate that, in the regime of
the medium-range order (between $r=4$\,\AA\,and $r=7$\,\AA), there are broad
peaks for Ni-Ni and Al-Al, which are either much larger or comparable in
magnitude to the primary peak associated with the short-range order (up to
$r\sim 3$\,\AA). In contrast to the La-pairs, in which the short-range order
peak appears to be the most dominant (Fig.~\ref{fig:rdf}a), the more active
Ni-Ni and Al-Al pair-interactions at the length-scale of the medium-range order
would also indicate a stronger dynamical coupling in this spatial regime.

\subsection{Vibrational density of states (VDOS)}
The filled gray circles in Fig.~\ref{fig:DOS} represent the total $D(\omega_p)$
as obtained from MD simulations. A more detailed look at the VDOS can be seen
through the respective contributions of the La, Ni and Al atoms. It is clear
that the initial maximum of the total $D(\omega_p)$ at around 8\,meV is
attributed to low-energy vibrations involving the heavy La atoms, while
vibrations of the Ni atoms occur around 15\,meV and are responsible for an
apparent shoulder on the high-energy side of the main vibrational band. The
vibrational dynamics of the light Al atoms are, in contrast, well separated
from that of the other elements and exhibit a double-band structure at around
25 and 35\,meV. The $D(\omega_p)$ as obtained in INS experiments is shown
alongside the simulation data. It is important to note here that the
experimental $D(\omega_p)$ is additionally weighted by the isotope-specific
neutron scattering cross-sections of the constituent elements, of which Ni-Ni
and Ni-La atomic pairs will dominate. Hence, the experimental $D(\omega_p)$
should be taken only to represent a generalized, neutron-weighted VDOS. In any
case, it is apparent that the predominant contribution to the high-frequency
side of both VDOS of this MG stems from the vibrations of the Al atoms.

\subsection{Dynamic mechanical analysis and comparison with theory}
In Fig.~\ref{fig:master} we show a master curve of the experimentally measured
$G''(\omega)$ obtained from Ref.~\cite{Zhu2014} for La$_{60}$Ni$_{15}$Al$_{25}$
at a reference temperature of 453 K, together with a theoretical fitting
provided by Eq. (5). The $\alpha$-relaxation appears as the main loss peak
situated around 1\,Hz. A distinct feature of this system is the prominent and
well separated loss peak on the high-frequency side around $10^6$\,Hz and is
attributed to the JG $\beta$-relaxation.

The nonaffine lattice dynamic theory of viscoelasticity of glasses
outlined above allows us to
quantitatively link the macroscopic features of the JG $\beta$-relaxation with
the atomic-scale vibrational properties of this MG}. Within this framework, it
is possible to rationalize the average friction in the atomic motion of a
tagged atom in the glass in terms of the respective contributions of the atomic
components, for which the friction coefficient of the $i$-th atom, $\nu_i$, is
proportional to the reciprocal of the atomic mass of atom
$i$~\cite{Zwanzig,Cui}. Thus, as was shown above in Sec. IV.B, when summing
over all tagged atoms in $i\omega\sum_{i}(\Phi_{ip})^2\tilde{\nu}_i$, the
contributions to the friction coefficient coming from the heaviest atoms, i.e.
La, turn out to be smaller by at least a factor of $1/3$ in comparison with the
contributions of Al and Ni (taken together). For the case of
La$_{60}$Ni$_{15}$Al$_{25}$ we thus find that the contribution of La can be
neglected, given the comparatively very large mass of La, which leaves the
average friction as the sum of two contributions, those of Ni and Al,
respectively, which carry widely different relaxation time scales, by virtue of
the different atomic masses.

\begin{figure}
\begin{center}
\includegraphics[height=5.7cm,width=8.6cm]{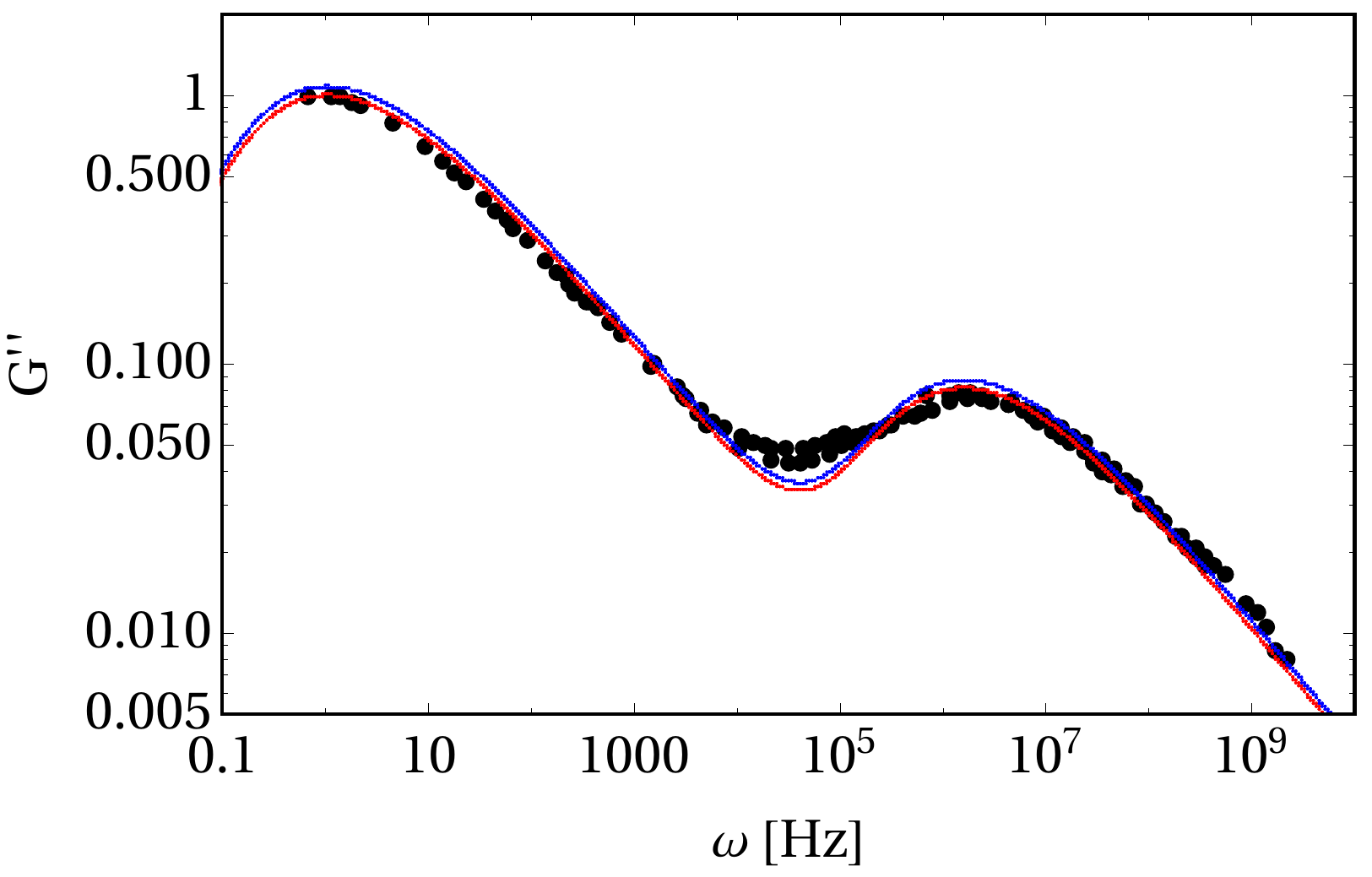}
\caption{\label{fig:master} Master curve of the imaginary part of the complex
viscoelastic modulus, $G''(\omega)$, at a reference temperature $T= 453$~\,K.
The red and blue curves are fit results to our theoretical model using the
experimental and simulated VDOS, respectively, as input.}
\end{center}
\end{figure}

As derived in Sec. IV.B, in the sum over $i$ only terms
corresponding to Ni and Al atoms survive, which are well separated in magnitude
given the difference in mass between Ni and Al. We then divide the sum into two
groups, for Ni and Al, respectively, and then average each group separately.
The final result is that the average friction memory function consists of two
distinct contributions, according to Eq. (7),
both of which will decay in time but with two different and well-separated
relaxation times, $\tau_1$ and $\tau_2$, respectively.
The shorter relaxation time $\tau_2$ (associated with the JG
$\beta$-relaxation) is related to the atomic dynamics of the lighter element,
Al, whereas the other term has a longer relaxation time $\tau_1$, dominated by
the atomic dynamics of the heavier element, Ni, which contributes to the
$\alpha$-relaxation time.

With an appropriate \textit{ansatz} for $\nu(t)$ we obtain the intermediate
scattering function $F(q,t)$ via $\nu(t)\sim F(q,t)^2$~\cite{Sjoegren,Bagchi}.
From experiments and simulations, we know that in supercooled liquids
$F(q,t)\sim \exp[(-t/\tau)^b]$ for the $\alpha$-relaxation, where $\tau$ is the
characteristic structural relaxation time and $b$ is the stretching exponent
with values normally between $b=0.5-0.7$~\cite{Hansen}.
When both $\alpha$- and $\beta$-relaxation are present, $F(q,t)$ has a two-step
decay, with a first decay at shorter times due to the $\beta$-relaxation, and a
second decay at much longer times due to the $\alpha$-relaxation. On the basis
of this evidence, we take the time dependence of each of the two terms in the
memory function to be stretched-exponential with different values of $\tau$ and
$b$,

\begin{equation}\label{eq:strexp}
\nu(t)\sim \exp[-(t/\tau_1)^{b_1}]+c\exp[-(t/\tau_2)^{b_2}],
\end{equation}

where $c$ is a constant.

The curves in Fig.~\ref{fig:master} are our fits to experimental data using the
VDOS obtained in both INS experiments (red) and MD simulations (blue). It is
apparent that our theoretical model excellently captures both peaks in the loss
spectrum over a frequency range of some 10 orders of magnitude with the
resulting parameters: $\tau_1=0.67$~\,s, $b_1=0.45$,
$\tau_2=4.04\cdot10^{-7}$~\,s, $b_2=0.47$ and $c=0.07$. We note here that the
two-component \textit{ansatz} is the simplest model with the minimum number of
free parameters that completely describes the experimental $G''$ data, which is
congruent with our theoretical result derived in the last section, where
$\nu(t)$ reduces
to a sum of two terms. Surprisingly, we obtain the same fitting parameters for
both the experimental and the simulation VDOS, although the two data sets
exhibit noticeably different features. In a way, this result reassures us that
the differences in the two VDOS didn't simply ``disappear" into the fitting
parameters and genuinely implies that these differences do not play a
substantial role in the mechanical response. Moreover, it suggests that the
qualitative shape of the VDOS, i.e. the location of the peaks, especially on
the low-frequency side, is of primary importance. In a broader perspective,
this result implies that the origin of the JG $\beta$-relaxation in various
types of glasses can be traced back to the generic shape of the VDOS and
encourages the development of a universal theory based on the microscopic
framework employed here.

\subsection{Qualitative behaviour of intermediate scattering function from
theoretical fitting}
\begin{figure}
\begin{center}
\includegraphics[width=\columnwidth]{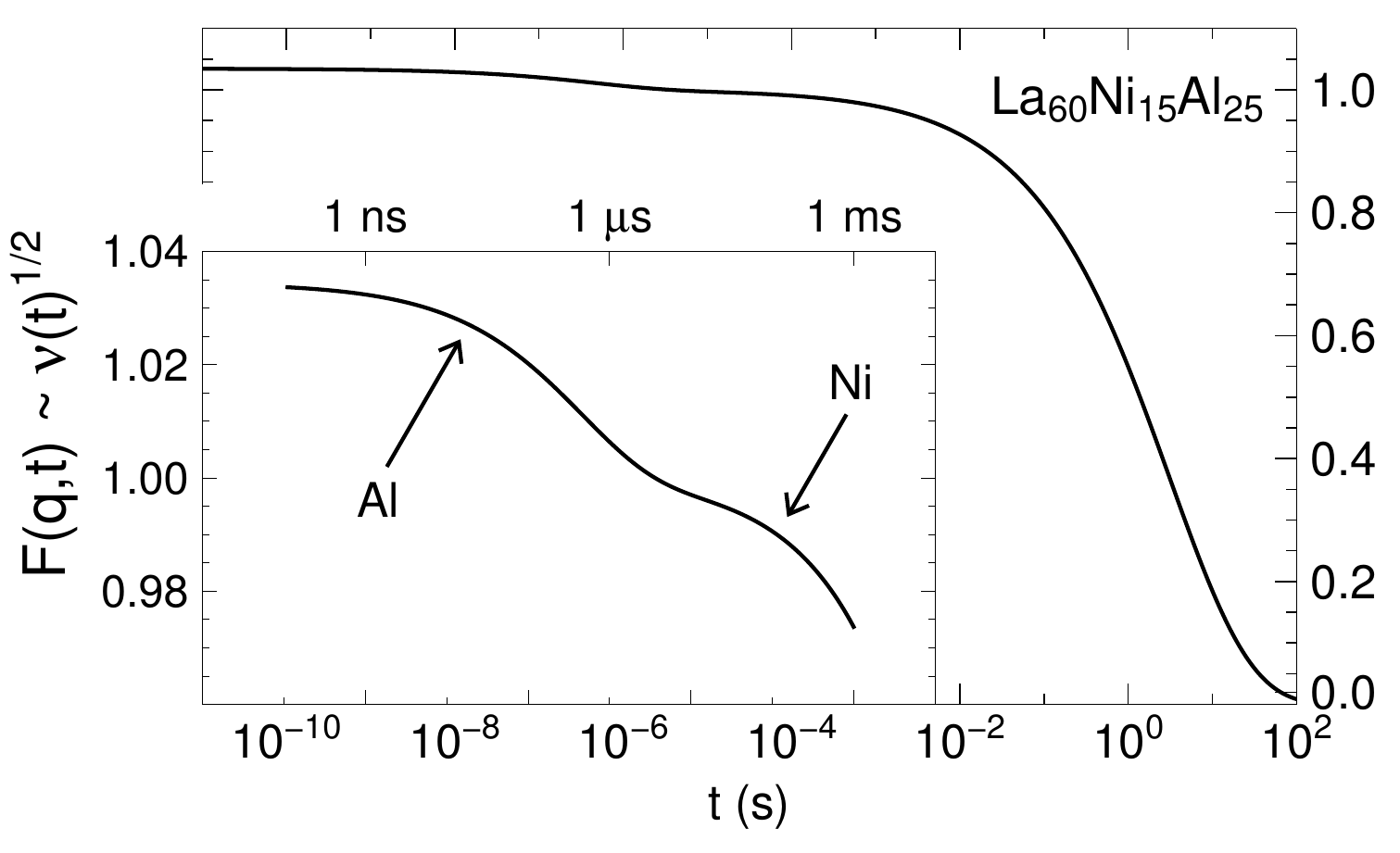}
\caption{\label{fig:ISF}Time decay of the square-root of total memory function
for the friction $\nu(t)$, exhibiting two decays corresponding to $\alpha$ and
$\beta$ decay in the intermediate scattering function $F(q,t)$, respectively,
according to the relation $F(q,t) \sim \sqrt{\nu(t)}$ that follows from
Eq.~(\ref{eq:Fqt})}
\end{center}
\end{figure}

The square-root of $\nu(t)$ is shown in Fig.~\ref{fig:ISF} following the
relation $F(q,t) \sim \sqrt{\nu(t)}$ from Eq.~(\ref{eq:Fqt}). We see the
characteristic two-step decay of $F(q,t)$ present in systems with well
separated $\alpha$ and $\beta$ relaxations, with the first decay occurring on
the typical time scale of the $\beta$ relaxation, $\tau_{\beta} \sim
10^{-7}~s$, followed by a much slower decay given by the time scale set by
$\tau_1$. While the time scale $\tau_{\beta}$ closely matches the time scale
$\tau_2$ set by atomic dynamics dominated by Al, the typical $\alpha$
relaxation time of glasses,  $\tau_{\alpha} \sim 10^{2}$~s, is significantly
different from the time scale $\tau_1$ associated with Ni, as the $\alpha$
process is more complex and the square-root mixing of the different time scales
of the above relaxation reflects this fact. Moreover, the $\alpha$ peak in
$G''$, and the corresponding decay in $F(q,t)$, cannot be reduced to just
$\tau_1$, as the time scale range of the $\alpha$-relaxation contains a strong
contribution from soft modes (the boson peak~\cite{Milkus}) in the VDOS. This
is clear from Eq.~(\ref{eq:GLE}) where the term $\omega_p^{2}$ in the
denominator
gives a large weight to the low-$\omega_p$ part of the VDOS, which contains the
BP-proliferation of soft modes, as was shown in previous work for the case of
CuZr alloys which present $\alpha$-relaxation only~\cite{Cui} and also for
dielectric relaxation of glycerol~\cite{Cui2}.

\section{Conclusion}
We have presented a combined experimental, simulation and theoretical analysis
of the viscoelastic response of a metallic glass exhibiting a strong
Johari-Goldstein (JG) $\beta$-relaxation. The appearance of the JG
$\beta$-relaxation in this metallic glass is attributed to (i) the wide mass
disparity between the light Al atoms and the other atomic species, and (ii) a
strong dynamical coupling involving the Ni and Al atoms at the medium-range
order length-scale. The results of our theory shed light
onto the microscopic glassy-state dynamics over a temporal range of 12 orders
of magnitude and reproduce the distinctive two-step decay of the intermediate
scattering function that is a characteristic feature of systems exhibiting both
$\beta$ and $\alpha$-relaxations. A crucial input to our theory is the
vibrational density of states (VDOS). Surprisingly, only the qualitative
features (i.e. peak positions) of the VDOS appear to play the main role in
determining the viscoelastic response of the glass, implying a common behavior
linking the JG $\beta$-relaxation to vibrational dynamics in glassy systems.
These results should be useful for developing a universal theory of
secondary relaxations in glasses.\\

\begin{acknowledgements}
We are grateful to the MLZ for the beamtime at TOFTOF. B. Cui acknowledges the
financial support from CSC-Cambridge Scholarship. P. Luo is gratefully acknowledged for sample preparation.
\end{acknowledgements}

\begin{appendix}

\section{Derivation of Eq. (4) in the main article}
After taking Fourier transformation of Eq. (3) in the main article, this
becomes
\begin{equation}
-\omega^2\tilde{\underline{x}}_i+i\tilde{\nu}_i(\omega)\omega\tilde{\underline{x}}_i
+\underline{\underline{H}}_{ij}\underline{\tilde{x}}_j
=\underline{\Xi}_{i,xy}\tilde{\eta}_{xy}.
\end{equation}

Next, we take normal mode decomposition. This is equivalent to diagonalize the
Hessian matrix $\underline{\underline{H}}$. From now on all matrices and
vectors are meant to be $3N \times 3N$ and $3N$-dimensional, respectively. The
$3N\times3N$ matrix
$\underline{\underline{H}}$ can be decomposed as
$\underline{\underline{H}}=\underline{\underline{\Phi}}~\underline{\underline{D}}~\underline{\underline{\Phi}}^{-1}=\underline{\underline{\Phi}}~\underline{\underline{D}}~\underline{\underline{\Phi}}^T$
where $\underline{\underline{D}}$ is a diagonal matrix filled with the
eigenvalues of $\underline{\underline{H}}$, that is, in components,
$D_{pp}=\omega_p^2$. Further, the matrix $\underline{\underline{\Phi}}$
consists of the $3N$ eigenvectors $\underline{\phi}_p$ of the Hessian, i.e.
$\underline{\underline{\Phi}}=(\underline{\phi}_1,...,\underline{\phi}_p,...,\underline{\phi}_{3N})$,
and is an orthogonal matrix.

Then, we left-multiply both sides with the matrix
$\underline{\underline{\Phi}}^{-1}=\underline{\underline{\Phi}}^T$, which leads
to Eq. (4) in the main article:
\[-\omega^2(\underline{\underline{\Phi}}^{T}\cdot\tilde{\underline{x}})+i\omega\underline{\underline{\Phi}}^T\tilde{\underline{\underline{\nu}}}(\omega)\underline{\underline{\Phi}}\underline{\underline{\Phi}}^{T}\cdot\tilde{\underline{x}}
+\underline{\underline{D}}~(\underline{\underline{\Phi}}^{T}\cdot\underline{\tilde{x}})
=\underline{\underline{\Phi}}^{T}\cdot\underline{\Xi}_{xy}\tilde{\eta}_{xy},
\]
where we used the fact that $\underline{\underline{D}}$ is diagonal and we have
dropped all indices $i$ and $j$ and $\tilde{\underline{\underline{\nu}}}$ is
the diagonal matrix $diag\{\tilde{\nu}_i\},i=1,2,...$.

\end{appendix}

\bibliographystyle{unsrt}

\end{document}